# The Nuclear $g_{9/2}$ Shell -Comparison of our work with an old B.H. Flowers Paper

Larry Zamick

Department of physics and astronomy Rutgers University, Piscataway, NJ 08854

## Abstract

In an old paper [1] B.H. Flowers discussed calculations for odd A nuclei in the $g_{9/2}$ shell. He finds when he varies the parameters of a certain interaction that the lowest energy state is always a seniority v=1 state with angular momentum J=9/2$^+$. However experiments show about half the states have J=9/2$^+$ lower and the other ones have J=7/2$^+$ lower. More recently we considered selected nuclei between Z=40 and Z=50. For lower odd N Zr isotopes we find J=7/2$^+$ comes below J=9/2$^+$ while for odd Z isotones near Z=50 the opposite is true i.e. J=9/2$^+$ comes below J= 7/2$^+$. We justify using different interactions in the 2 regions by considering nuclear deformation near the Z=40 region. A plea to find missing energy levels is made.

1. Discussion

A slight paraphrase of the abstract of the paper of B.H. Flowers [1] I here shown:

"Wave functions and energy level diagrams for 3 identical particles in the $g_{9/2}$ are calculated. The J=7/2$^+$ level of seniority three is shown never to be the ground state; this is in conflict with experimental findings and suggests a breakdown of the jj coupling approximation. "

Another slight paraphrase on the experimental situation:

"Now Goldhaber and Sunyar [2] report in about half the cases in which the J=9/2 (s=1) and J=7/2 (s=3) ( of the $g_{9/2}^n$ configuration for n=3,5/,7) can be identified about half the J=7/2 level lies lower than the J=9/2 level."

Flowers used an interaction V(r)=const. *exp-(r/a)$^2$ . He notes the $g_{9/2}$ radial wave function is of the form
constant r^4 exp(-1/2 (r/a$_0$)$^2$) . He plots results as a function of the dimensionless parameter (a/a0)$^2$ . He found that for $g_{9/2}^3$ no matter what value this parameter assumes the J=9/ 2$^+$ level came out lower than J=7/2 $^+$ .

In more recent work by Escuderos and Zamick [3] much attention was given to their finding that for identical particles of the $g_{9/2}^4$ configuration there was a unique J=4 state with seniority 4;likewise J=6. This was a surprise because seniority was supposed not to be a good quantum number in the $g_{9/2}$ shell. Only in $f_{7/2}$ and below was seniority a good quantum number.

There is a second part of the paper that deals with $g_{9/2}^n$ for odd n. This has not received as much attention but there are some good points to be made.

2. The $g_{9/2}$ Model Space

In the work of Escuderos and Zamick [3] they consider nuclei consisting of proton holes relative to 100Sn and neutrons relative to $^{80}$Zr. We call this the $g_{9/2}$ model space rather than the $g_{9/2}$ shell because it is known the 2 extremes above are quite different. Most researchers including the present author are quite comfortable with employing the shell model for proton holes relative to $^{100}$Sn. But as was shown in Hartree-Fock calculations [4], $^{80}$ Zr is highly deformed-beta is around 0.4. This was reinforced by Zheng and Zamick [5] who also considered excited n- particle-n-hole bands. This will play a key in explaining the observation by Flowers [1] that not all odd A in this region have J= 9/2$^+$ states. Loosely speaking we can say that $^{100}$Sn is a good closed shell whilst $^{80}$Zr is a bad one.

In the work of Escuderos and Zamick [3] single j ($g_{9/2}$) calculations are performed for the following nuclei :

$^{97}$Ag - 3 proton holes

$^{95}$Rh – 5 proton holes

$^{93}$Tc - 7 proton holes or 3 protons.

$^{83}$Zr - 3 neutrons

$^{85}$Zr – 5 neutrons

$^{87}$Zr – 7 neutrons or 3 neutron holes

We first discuss experiment . We refer to Fig1 of ref [3.]

For $^{97}$Ag, $^{95}$Rh and $^{93}$Tc the ground state has J=9/2 $^+$. For the latter 2 nuclei the J=7/2$^+$ states are over half an MeV above. Surprisingly the J= 7/2$^+$ state has not been identified in $^{97}$Ag, even to this day. But we are sure it is not the ground state.

In the Zr isotopes we must mention that there are 2 negative parity states below the positive parity ones- J=1/2$^-$ and 5/2$^-$ .This underscores the fact that $^{80}$Zr is not a good closed shell.

In $^{83}$Zr and $^{85}$Zr the J=7/2$^+$ state lies below J=9/2$^+$ whilst in. $^{87}$Zr J=9/2$^+$ is lower.

In a shell model calculation with a fixed charge independent interaction and limited to a single j shell the spectra for n proton holes relative to $^{100}$Sn would be the same as that of n neutrons relative to $^{80}$Zr. In order to get different spectra one must use different interactions for the 2 cases.

Near $^{100}$Sn one can get the 2 body interaction matrix elements from experiment i.e. the spectrum of 98 Cd -a 2 proton hole system. We call this V($^{98}$Cd). In Fig 4,5 and 6 get the robust result that with this interaction J=9/2$^+$ is calculated to be lower than J=7/2$^+$ for $^{97}$Cd, $^{95}$Rh and $^{93}$Tc, in agreement with experiment.

As just mentioned if one used V($^{98}$Cd) for Zr isotopes we would get the spectra as for the above nuclei. To explain the Zr region we. Must use a drastically different interaction.

In ref [3] we noted that the quadrupole-quadrupole interaction Q.Q yields the result that J=7/2$^+$ comes below J=9/2$^+$. One can use a linear combination of Q.Q and a delta interaction to fit all nuclei ,those close to $^{100}$Sn and those close to $^{80}$Zr. In the $^{100}$Sn region one has much more delta whist in the $^{80}$Zr region one needs much more Q.Q. The latter seems reasonable -in a deformed region Q.Q dominates over delta.

Another point of interest from Ref [3] is the energy splitting E( 21/2$^+$) – E(3/2$^+$) for n=3 and 5. Here J=21/2 is the highest angular momentum for $(g_{9/2})^3$ and J= 3/2 is the lowest. With a seniority conserving interaction this splitting is the same for n=3 and n=5. With a Q.Q interaction the splitting is equal but opposite for n=3 and n=5. Our calculations show that this splitting should be positive in $^{97}$Cd and $^{95}$Rh but positive for $^{83}$Zr and negative for $^{85}$Zr.

Unfortunately no J=3/2$^+$ states have been identified experimentally for any of the nuclei in mentioned here.. We hope this work will stimulate a search for such states.

Although the Q.Q interaction correctly gives the J=7/2 J=9/2 splitting for $^{83}$Zr it is too much to ask a $(g_{9/2})^3$ to .give a detailed spectrum. Other approaches such as that of Huttmeier et al. [6] are required.

Besides references to the second part of [3] we add those to the first part following ref[6].

Closing remarks:

Flowers [1] put his finger on an important problem-getting the right level ordering in the $g_{9/2}$ shell. He in particular focused on the J=9/2$^+$ and J=7/2$^+$ levels in odd A nuclei. We find that to solve this requires fairly drastic measures. This is mainly due to the fact that $^{100}$Sn is spherical but $^{80}$Zr is highly deformed. In $g_{9/2}$ model space one needs quite different interactions in the A=80 region than what one uses in the A=100 region.

We also hope that a surprising number of missing levels be found experimentally in the near future. This will help to sharpen our analysis of this region of the periodic table.

**References**

[1] B.H. Flowers, Proc. Royal Society of London, Series A,, Vol 215,1122,==398-403 (1952 )

[2] M. Goldhaber and A.W .Sunyar, Phys. Rev. 83,906 (1951)

[3] A. Escuderos and L.Zamick, Physical Review C 73 ,044302 (2006 )

[4] P.Bonche, H.Floccard, P.H. Heenan, S.J. Kreiger and M.S.Weiss, Nucl. Phys A443,39 (1985)

[5] D.C. Zheng and L. Zamick, Physics Letters B266,5(1991)

[6] U.J. Huttmeier , C. J. Gross, D.M. Headly, E.F. Moore, S.L. Tabor, T.L. Cormier, P.M. Stewartka and W.Nazarewicz,Phys. Rev. C37, 118 (1988)

[7] G.Racah,Phys.Rev.62,438(1942);63,37(1943);76,1352(1949)

[8] A. De Shalit and I. Talmi ,Nuclear Shell Theory,Academic Press,New York (1963)

[9] I.Talmi, Simple Models of Complex Nuclei ,Harwood Academic Press,Switzerland (1993)

[10] B.F.Bayman and A.Lande, Nucl. Phys. 77,1 (1966)

[11] L. Zamick, Phys. Rev. C75, 064305 (2007)

[12] L. Zamick and P. Van Isacker, Phys. Rev. C78, 044327 (2008)

[13] I. Talmi, Nucl. Phys. A846, 31 (2010)

[14] P. Van Isacker and S. Heinze, PRL 100, 052201 (2008)

[15] P. Van Isacker and S. Heinze, Annals of Physics 348, 73 (2014)

[16] C. Qi, Phys. Rev. C81, 014307 (2011)

[17] C. Qi, Z.X. Xu. R.J. Liotta, Nucl. Phys. A 844, 855 (2012)

[18] X.B. Wang and F.R. Xu, Phys. Rev. C85, 03404 (2012)

[19] C. Qi, Phys. Lett., B773, 616 (2017)

[20] P.C. Srivastava, Bharti Bhoy and M.J. Emamtov, Progress of Theoretical and Experimental Physics 2019, 103D01 (2019), doi 10.1093/ptep/plz108

[21] A.J. Morales et al., Physics Letters B71, 206 (2018)

[22] A.T. Krupa, J. Kovacs, P. Salamon, O. Legeza, and G. Zarand, Phys. Rev. C106, 024303 (2022)

[23] B. Das et al., Phys. Rev. C105, L031304 (2022)

[24] K. Neergard, Phys. Rev. C106, 024308 (2022)

[25] R.M. Perez-Vidal et al., Phys. Rev. Lett., 129, 112501 (2022)

[21] S. Siderov, D.Zhulyaeva and T.Tretyakova, Chinese Physics C ,00001 (2022)